\documentclass[twocolumn,showpacs,preprintnumbers,amsmath,amssymb]{revtex4}

\usepackage{graphicx}
\usepackage{natbib}

\begin{document}

\title{An experimental test of volume-equilibration between granular systems}
\author{Fr\'ed\'eric Lechenault, Karen E. Daniels}
\email{kdaniel@ncsu.edu}
\affiliation{Department of Physics, Box 8202, 
North Carolina State University, Raleigh, NC, USA 27695}

\date{\today}

\begin{abstract}
Understanding granular and other athermal systems requires the identification of state variables which consistently predict their bulk properties. A promising approach has been to draw on the techniques of equilibrium statistical mechanics, but to consider alternate conserved quantities in place of energy. The Edwards ensemble \citep{Edwards-1989-TP}, based on volume conservation, provides a temperature-like intensive parameter called compactivity. We present experiments which demonstrate the failure of compactivity to equilibrate (via volume-exchange) between a pair of externally-agitated granular subsystems with different material properties. Nonetheless, we identify a material-independent relationship between the mean and fluctuations of the local packing fraction which forms the basis for an equation of state. This relationship defines
an intensive parameter that decouples from the volume statistics.
\end{abstract}

\maketitle

Granular materials such as sand exhibit a transition from solid-like to liquid-like states which cannot be predicted from classical thermodynamics due to the separation of energy scales: thermal fluctuations are insufficient to cause grain rearrangements. However, because the behaviors of granular materials are strongly reminiscent of ordinary phases, there have been extensive efforts to formulate statistical theories as explanations. The Edwards proposal \citep{Edwards-1989-TP} to formulate a volume-based (as opposed to energy-based) ensemble of states has provided the basis for much work in the field \citep{Nowak-1998-DFV, Barrat-2000-EMP, Brujic-2003-JSS, Schroter-2005-SSV, Lechenault-2006-FVD, Blumenfeld-2006-GPF, Makse, Aste}. Within this approach, a temperature-like parameter called  compactivity plays the role of the central state variable, yet fundamental properties such as equilibration have not been established. 

Following the Edwards approach, several ensembles have been proposed in order to define the associated microcanonical entropy. In general, these ensembles are grounded in partition functions relying on grain-scale spatial tessellations \citep{Edwards-1989-TP, Blumenfeld-2006-GPF, Makse} which have recently been generalized to include stresses \citep{Edwards-2005-FCE, Henkes-2007-ETS, Metzger-2008-HTC, Tighe-2008-EMF, Brujic-2003-JSS}. In each case, the temperature-like intensive variable appears in the denominator of a Boltzmann-like weight and should take the same value in subsystems allowed to exchange the conserved quantity. For the Edwards ensemble, the microcanonical entropy is $S(V) = k \log{\Omega}$ where $\Omega$ is the number of mechanically stable configurations compatible with the overall volume $V$. The associated intensive parameter $X$ is known as the compactivity and is defined, by analogy with thermodynamics, as $\frac{1}{X} = \frac{\partial S}{\partial V}$. The compactivity measures how far the system is dilated above its most compact state ($X = 0$). While several measurements of the compactivity have been performed in experiments \citep{Nowak-1998-DFV, Schroter-2005-SSV, Lechenault-2006-FVD, Aste}, none has yet assessed equilibration between subsystems. 

In this paper, we present experiments which demonstrate that compactivity does not equilibrate between granular subsystems exchanging volume and kept in a dynamical stationary state. The two subsystems have the same number of grains and differ only in their material properties, but are observed to occupy different total volumes. Encouragingly, we find a species-independent, one-to-one relationship between the mean local packing fraction $\phi$ and its fluctuations. Together, these two results provide different values of the compactivity for the two subsystems. While the observed constant ratio between $\langle \phi \rangle$ and its fluctuations provides a well-defined intensive parameter, it cannot provide a prediction for the value of the two volumes. This points to the need for an ensemble that captures additional aspects of the granular material close to its jamming transition, such as the coupling between volume and internal forces and/or dynamical features. Such a broader theory must provide other intensive quantities which do equilibrate, allowing for the establishment of the full equation of state.

\begin{figure}
\centerline{\includegraphics[width=0.45\textwidth]{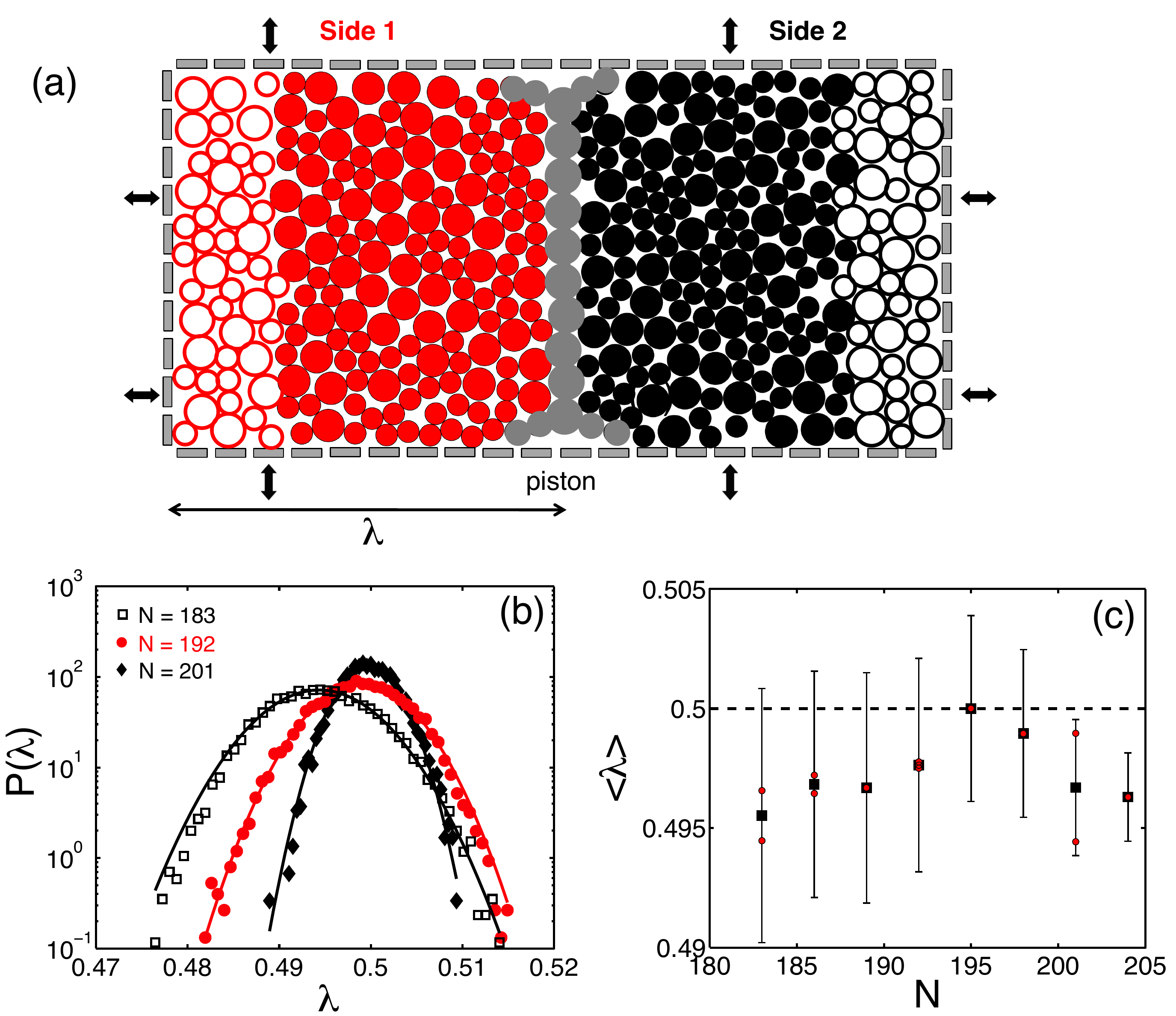}}
\caption{(a) Apparatus schematic (1 m $\times$ 2 m), with particles drawn to scale. Filled grains are taken from an experimental configuration and indicate the size of the imaged region. $\lambda$ measures the fractional position of the piston. (b) Sample probability density functions of $\lambda$ (and Gaussian fits) for $N=183$, $192$ and $201$ ($\bar{\phi}$ = 0.766, 0.789, 0.812). (c) $\langle \lambda \rangle$ for Side 1 as a function of $N$, with bars representing the standard deviation. For some values of $N$, the experiment has been repeated several times starting from different initial configurations, in order to evaluate the errors; these measurements are shown as smaller symbols.}
\label{fig:schematic}
\end{figure}

We study the equilibration of two adjacent dense bidisperse layers of disks on an air table. (see Fig.~\ref{fig:schematic}). A piston separates the two subsystems, and is constrained to move only along the long axis of the air table. On each side, the system is prepared with an identical number of grains $N$, with the ratio of small to large grains fixed at $N_S = 2N_L$. This ratio suppresses crystallization and provides roughly the same volume of large (diameter $d_L = 86$ mm) and small grains ($d_S = 58$ mm). The disks within each subsystem have a different set of material properties: on Side 1, the particles have restitution coefficient $\epsilon_1 = 0.51 \pm 0.07$ and friction coefficient $\mu_1 = 0.85$; on Side 2, the particles have $\epsilon_2 = 0.33 \pm 0.03$ and $\mu_2 = 0.5$. Both sides utilize standard plastic Petri dishes as the particles, with the difference in particle properties achieved by encircling the particles on Side 1 with a rubber band. Restitution coefficients were measured from isolated binary collisions; friction coefficients are nominal values from the literature. Because the sides of the particles slope inwards, the thickness of the rubber band does not significantly change the radius of the particles; the mass of the particles on Side 1 is increased by 7\%. 

The aggregate rearranges via an array of sixty electromagnetic bumpers which form the walls of the system. These bumpers are triggered pairwise: bumpers facing each other in the system fire at the same time in order to prevent net momentum and torque injection. 
Four pairs of bumpers are randomly fired every $0.1$ second, and travel 1 cm into the granular pack;
the total time during which the bumpers stay in the ``active'' position is $\approx 0.095$ second.

To quantify the long-time mobility of the particles, we take images at a frequency which is low compared to the energy injection timescale, making usual tracking techniques inoperative. Therefore, we have developed a tracking method which identifies each particle by a unique tag. Each particle is marked with a $3 \times 3$ array of colored dots which encodes two copies of a 4-bit, 4-digit identifier, plus an error-correcting bit. The particles are located by their circular rims and their identities are established using the tags, allowing adjacent image frames to be connected into trajectories. We monitor the positions of the piston and the inner 75\% of the disks with a CCD camera mounted above the apparatus; we obtain a minimum of $10^4$ configurations for each experiment. 

\begin{figure}
\centerline{\includegraphics[width=0.4\textwidth]{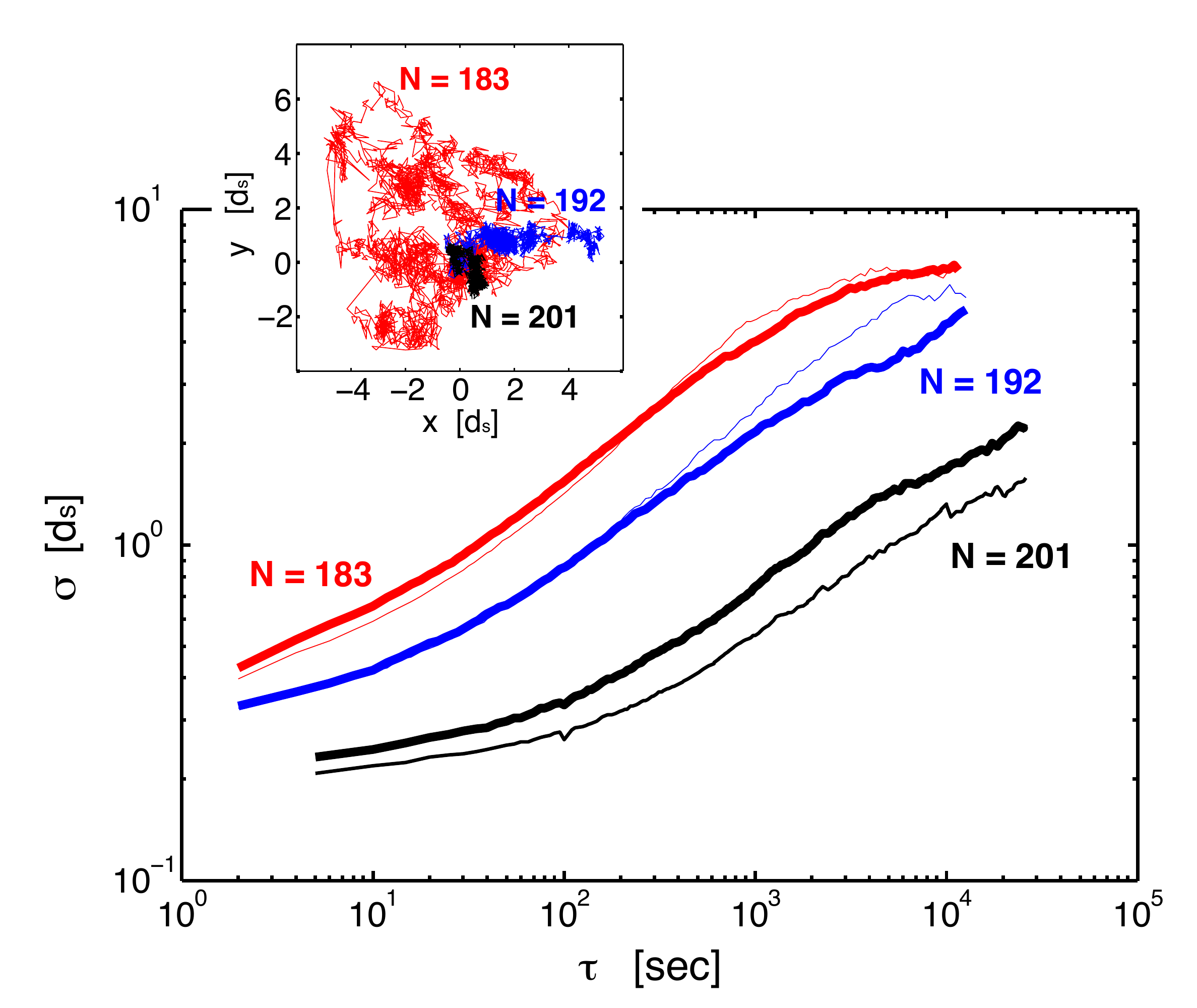}}
\caption{Diffusion length $\sigma$ as a function of lag-time $\tau$. Thin lines are for Side 1; thick lines are for Side 2. Inset: Trajectories of a single disk (from Side 1) for a duration of $10^4$ sec.}
\label{fig:diffusion}
\end{figure}

To understand the equilibration of the system on its approach to jamming, we perform experiments at increasing values of $N$ while holding all other variables constant. Over the studied range $N=183$ to $204$ (corresponding to $\bar{\phi}\equiv 2\left(\frac{1}{\langle \phi_1 \rangle} + \frac{1}{\langle \phi_2 \rangle}\right)^{-1} = 0.768$ to $0.818$), the system transitions from a liquid-like to a solid-like state. To monitor this transition, we use particle-trajectories to compute the average diffusion length, defined as
$\sigma \left(\tau \right) \equiv \sqrt{\langle\|\vec{r}_i\left(t+\tau\right) - \vec{r}_i\left(t\right)\|^2\rangle_{i,t}}$
where $\vec{r}_i\left(t\right)$ is the vector position of disk $i$ at time $t$. Fig.~\ref{fig:diffusion} shows a plot of this quantity as a function of $\tau$ for three values of $N$ and for each side. The inset shows typical trajectories at corresponding values of $N$. At $N=183$ (low $\bar \phi$) a particle explores a region several particle diameters wide, and $\sigma$ saturates at long time scales due to finite system size. On the other hand, at $N=201$ (high $\bar \phi$), $\sigma$ exhibits a subdiffusive plateau at short $\tau$, and caging effects are significant, as can be seen in Fig.~\ref{fig:diffusion}. This change in mobility occurs as $\phi$ approaches and crosses $\phi_{RLP}$. Therefore, as the dynamics of the system slows down, we increase the duration of the experiment from twenty hours up to fifty-five hours for the densest packing in order to sample a significant set of configurations.

The scaled position of the piston, $0 < \lambda < 1$, measures the macroscopic state of each subsystem. After a transient, the probability distribution of $\lambda$ becomes stationary and Gaussian, even for the largest $N$; sample distributions are shown in Fig.~\ref{fig:schematic}c. We examine $\langle \lambda \rangle$ as a function of $N$ and observe that $\langle \lambda \rangle \leq \frac{1}{2}$. On average and for all values of $N$, Side 1 occupies less than half of the overall volume.  We conclude that this systematic deviation from equal volumes originates from the difference in material properties of the disks. Therefore, the density of states depends not only on the available space, but also on the grains' properties.

\begin{figure}
\centerline{\includegraphics[width=0.49\textwidth]{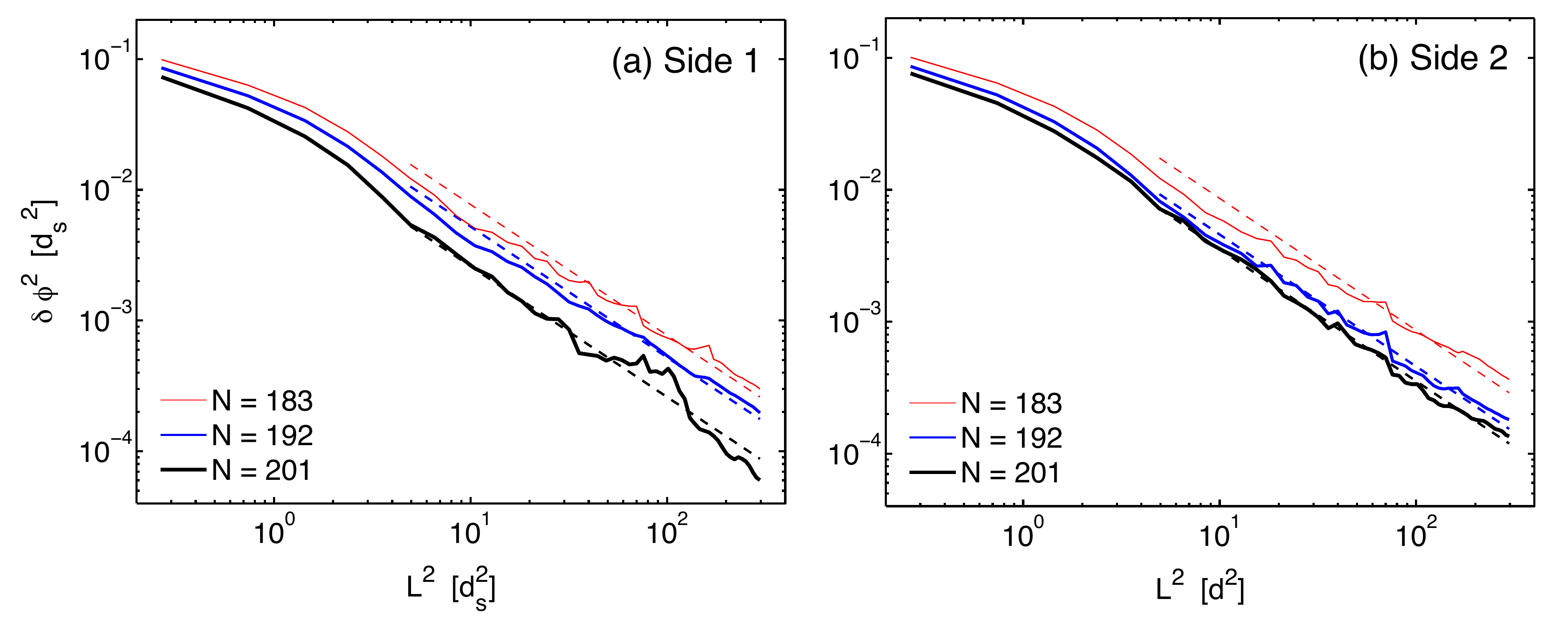}}
\caption{Variance $\langle \delta \phi^2 \rangle$ of the packing fraction measured within squares of size $L$ for three values of $N$; dashed lines are $\langle\delta\phi^2\rangle_0 / L^2$. }
\label{fig:dphi}
\end{figure}

In order to extract a microscopic, ``canonical'' quantity, we analyze the temporal fluctuations of the average local $\phi$ over windows of increasing size. For each subsystem, we measure $\phi$ over boxes of size $L$ ranging from a few particle diameters up to half the system size. We observe that the variance $\langle \delta \phi^2 \rangle$ scales approximately as $L^{-2}$, as can be seen on Fig.~\ref{fig:dphi}. Such scaling behavior is expected from the central limit theorem in the absence of long-ranged spatial correlations in $\phi$. We can therefore conclude that the packing fraction is a self-averaging quantity, with statistics suitable for a thermodynamic-like analysis, despite the relatively small number of particles. We obtain an $L$-independent measure of the variance of $\phi$ by averaging $\langle \delta\phi^2 \rangle _0 \equiv \langle \delta \phi^2 \rangle  L^2 $ over all $L > 2 d_S$.

\begin{figure}
\centerline{\includegraphics[width=0.45\textwidth]{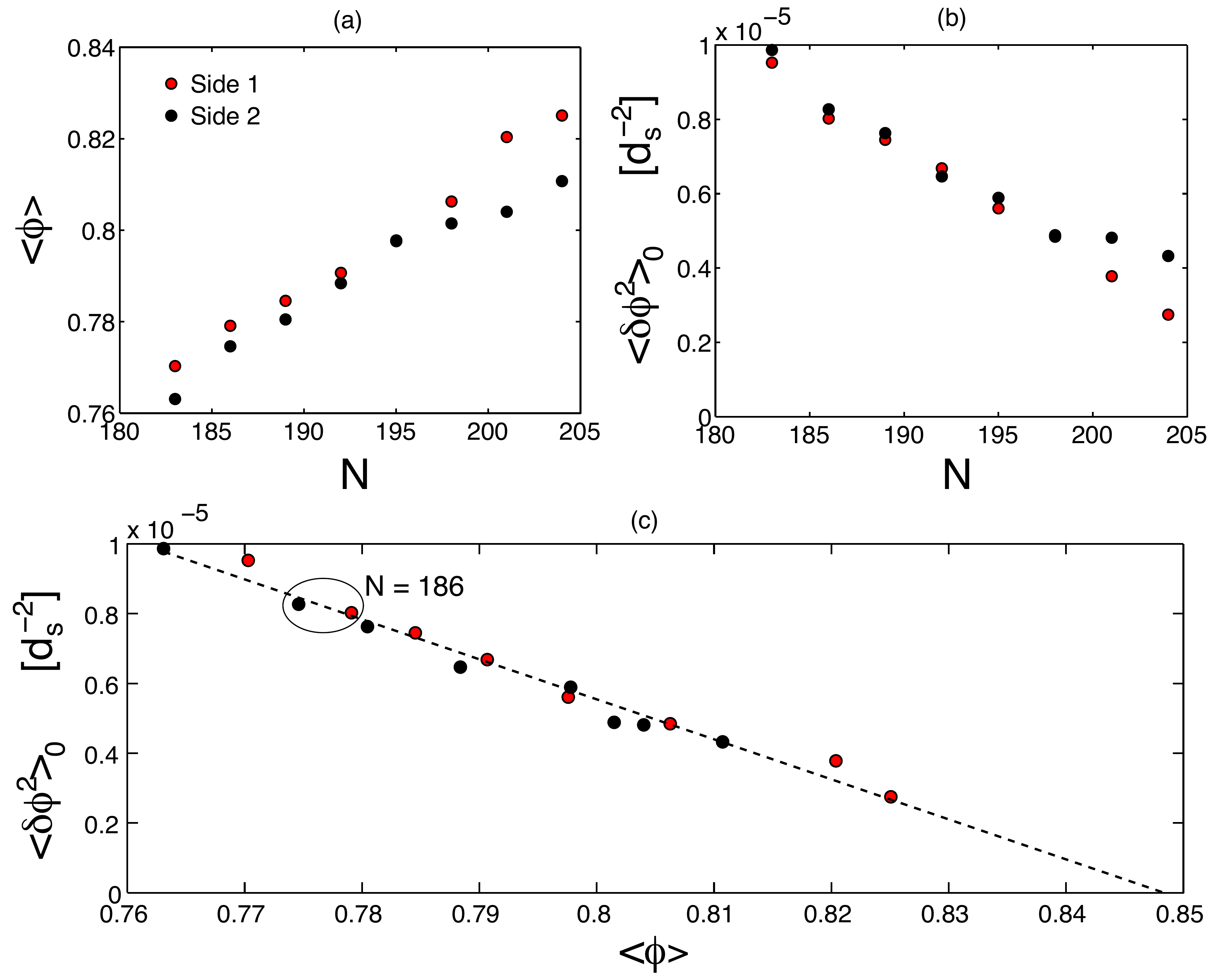}}
\caption{(a) Packing fraction $\langle \phi \rangle$ and (b) normalized packing fraction fluctuations  $\langle \delta\phi^2 \rangle_0$ as a function of number of particles on each side. (c) Data from parts (a) and (b) combined on a single plot without regard for $N$. Ellipse encloses two points obtained from Side 1 (red) and Side 2 (black) for a single run with $N=186$. Dashed line is a linear fit to data from both sides, showing an intercept at $\phi = 0.848$.}
\label{fig:equi}
\end{figure}

This analysis provides both the mean packing fraction $\langle \phi \rangle$ and its normalized fluctuations $\langle \delta\phi^2\rangle_0$ defined as $\langle\delta\phi^2\rangle(L) = \frac{\langle \delta\phi^2\rangle_0}{L^2}$. These packing fraction statisitics are plotted as a function of $N$ in Fig.~\ref{fig:equi}a,b. 
The behavior of the two sides reproduces our macroscopic measurement: $\langle \phi \rangle$ is higher for Side 1. At $N=195$ the packing fractions are equal ($\langle \phi \rangle = \bar{\phi} = 0.798$ for both sides), as shown in Fig.~\ref{fig:equi}a. This value is in approximate agreement with independent measurements of random loose packing for these grains: $\phi^{RLP}_1 = 0.807 \pm 0.010$ and $\phi^{RLP}_2 = 0.812 \pm 0.006$. To determine $\phi^{RLP}$ for each of the two types of particles, we placed the table at a $0.3^\circ$ angle, and rained down single particles from the high end to the low end to create the loosest packing accessible to us. In addition, this value of $N$ corresponds to a transition from Brownian to caged dynamics (see Fig.~\ref{fig:diffusion}).

Different values of the normalized $\phi$ fluctuations $\langle \delta\phi^2\rangle_0$ are observed on the two equilibrated sides, indicating that the statistics of this quantity depend on the material properties. Moreover, the fluctuations on each side are found to be decreasing functions of $N$, as shown in Fig.~\ref{fig:equi}b. This can be understood by the fact that as $N$ increases, the amount of free volume to be distributed among the grains decreases.
Finally, the discrepancy in the packing fraction statistics between the two sides gets larger on approach to jamming at large $\bar \phi$ (large $N$).

To examine the relevance of these state variables, we plot the dependence of $\langle \delta\phi^2\rangle_0$ on $\langle \phi \rangle$ (see Fig.~\ref{fig:equi}c). Remarkably, the data from both sides fall on a master curve: the fluctuations in $\phi$ are uniquely determined by the local free volume, and are {\itshape insensitive} to the material properties of the grains. Within the explored range of $N$, this master curve is approximately linear and the extrapolation would intersect $\langle\delta\phi^2\rangle_0=0$ at $\langle \phi \rangle = 0.845$, which is compatible with the rigidity transition reported in other bidimensional systems \citep{OHern-2003-JZT, Majmudar-2007-JTG, Lechenault-2008-CSH}.

In order to measure the compactivity, we follow the method of \citet{Nowak-1998-DFV} We will consider that each side is a compactivity reference, or {\it compactostat}, and use the microscopic information as a {\it compactometer}. Within this canonical framework, we start from the Edwards partition function 
${\cal Z} = \sum_{ \cal C}  e^{-V({\cal C})/X}$
where the summation runs over the set ${\cal C}$ of mechanically stable configurations of the assembly, and $X$ is the compactivity. In our system, configurations above $\phi^{RLP}$ are mechanically stable and thus the ensemble explored by the rearrangements closely matches the Edwards ensemble. By analogy with the standard canonical ensemble, successive derivatives of $\ln {\cal Z}$ provide higher moments of the volume distribution. In particular, the mean volume and its fluctuations are related by
\begin{equation}
\langle \Delta V^2 \rangle = X^2 \frac{d \langle V \rangle}{dX} 
\label{e:Vfluct}
\end{equation}
We integrate Eq.~\ref{e:Vfluct} from a reference state with known values $(X_{\mathrm{ref}}, V_{\mathrm{ref}})$ to the current state $(X, V)$, yielding the relation
\begin{equation}
\frac{1}{X_{\mathrm{ref}}} - \frac{1}{X} =  \int_{V_{\mathrm{ref}}}^{V} 
\frac{dV}{\langle \Delta V^2 \rangle} 
\label{e:Xint} 
\end{equation}
In our experiments, the $\phi$-statistics are observed to be identical for the two subsystems at $N=195$. Therefore, we can use this as a reference state with $X_{\mathrm{ref}}$ and $V_{\mathrm{ref}}$ equal for both sides. In Fig.~\ref{fig:equi}c, we observe that the integrand in Eq.~\ref{e:Xint} is the same for both sides. Therefore, we can perform the integral from $N_{\mathrm{ref}} = 195$ to any value of $N$ and obtain the compactivity measured relative to $X_{\mathrm{ref}}$. We are particularly interested in the jammed states ($N > 195$): performing the integral over the same integrand to two different values of $V$ yields two different values of $X$. Therefore, $X_1 \ne X_2$ even though the subsystems are equilibrated (in steady state). 

This conclusion relies on the unexpected one-to-one relationship between the average packing fraction and its local fluctuations. This relationship should be predicted from an as-yet unknown equation of state. It is tempting to define a variable $Y\equiv \frac{\langle \delta\phi^2\rangle_0}{\langle\phi\rangle}$ which is intensive, $\phi$-independent, and takes the same value in both (all) subsystems. However, the value of $Y$, which we conjecture to be related to the energy injection and/or dissipation rates in the system, is not enough to specify the equilibrated volumes. This suggests that another parameter is needed to fully characterize the state of a dynamically evolving dense granular pack, much as both temperature and pressure equilibration are needed to solve the equivalent problem within classical thermodynamics.

It is instructive to speculate about the nature of the missing information. First, the Edwards ensemble relies on an assumption that configurations occupying the same volume are equiprobable. Recent simulations \citep{Gao-2006-FDM} on frictionless particles have challenged the validity of this assumption. However, even for unequal probabilities, the integral in Eq.~\ref{e:Xint} remains valid \citep{Lechenault-2006-FVD}, and thus the measured compactivity would still fail to equilibrate. Second, higher $\mu$ or lower $\epsilon$ might provide for more efficient energy dissipation in the bulk and couple the grains differently to the energy injection at the boundaries. Measuring energy transfer rates might prove useful to decipher the statistics of such a system \citep{Bonetto-2006-FRE}, and studies of grain kinetics would allow further investigation of such a mechanism. An approach based on a fluctuation/response ratio \citep{Potiguar-2006-ETJ} provides an additional notion of effective temperature closely related to these dynamical features. Finally, and importantly, intermittent force chains are particularly abundant in higher-density states \citep{Howell-1999-SF2} and likely couple the microscopic stress state to the macroscopic volume. Recent proposals \citep{Edwards-2005-FCE, Henkes-2007-ETS, Metzger-2008-HTC, Tighe-2008-EMF} to include contact force or stress statistics in defining intensive statistical parameters might provide the solution. 

These experiments have examined microscopic and macroscopic jammed matter statistics in light of existing theories. Remarkably, a grain-independent relationship was found between packing fraction average and fluctuations which provides a well-defined state variable, although insufficient to fully characterize the state of the dynamic packings; further investigations are needed in order to establish the generality of this observation. Our measurements also highlight the important role played by the static $\phi^{RLP}$ in the dynamics of a granular system. The connection between $\phi^{RLP}$, the  jamming point $\phi^{J}$, and granular glassy dynamics \citep{Marty-2005-SCE} should be further investigated. Finally, we observed that subsystems with different material properties do not in general share volume equally. This underlies the crucial role played material properties such as friction in determining the state of a granular packing. 

{\bfseries Acknowledgments:} The authors thank R. Blumenfeld, J. P. Bouchaud, O. Dauchot, B. Chakraborty, S. F. Edwards, R. P. Behringer, P. Reis, M. Schr\"oter, M. Shearer, and B. P. Tighe for informative discussions during the development and analysis of these experiments. We are grateful for support from the NSF under grant DMR-0644743.



\end{document}